\documentclass[aps,prl,twocolumn]{revtex4}
\usepackage{graphicx} 
\usepackage{amsmath}
\usepackage{amsfonts}
\usepackage{amssymb}
\usepackage{gensymb}
\usepackage{color}

\newcommand{\ket}[1]{|#1\rangle}
\newcommand{\bra}[1]{\langle #1|}

\begin{document}
\title{Monopole field textures in interacting spin systems}
\author{Andreas Eriksson}
\affiliation{Department of Physics and Astronomy, Uppsala University, Box 516,
Se-751 20 Uppsala, Sweden}
\author{Erik Sj\"oqvist}
\email{erik.sjoqvist@physics.uu.se}
\affiliation{Department of Physics and Astronomy, Uppsala University, Box 516,
Se-751 20 Uppsala, Sweden}
\date{\today} 
\begin{abstract}
Magnetic monopoles can appear as emergent structures in a wide range of physical settings, 
ranging from spin ice to Weyl points in semimetals. Here, a distribution of synthetic (Berry) 
monopoles in parameter space of a slowly changing external magnetic field is demonstrated 
in a system of interacting spin-$\frac{1}{2}$ particles with broken spherical symmetry. These 
monopoles can be found at points where the external field is nonzero. The spin-spin interaction 
provides a mechanism for splitting the synthetic local magnetic charges until their magnitude 
reaches the smallest allowed value $\frac{1}{2}$. For  certain states, a nonzero net charge can 
be created in an arbitrarily large finite region of parameter space. The monopole field textures 
contain non-monopolar contributions in the presence of spin-spin interaction.  
\end{abstract}
\maketitle
While magnetic monopoles seem up to this date mysteriously absent as fundamental entities 
in nature, they may occur as emergent structures in various physical systems. Indeed, real 
space realizations  of such emergent monopoles have been demonstrated in spin 
ice \cite{castelnovo08} and Bose-Einstein condensates \cite{ray15}, but also in reciprocal 
space of crystalline systems, e.g., in the context of anomalous Hall effect \cite{fang03}, as 
well as in the form of Weyl points in semimetals \cite{lv15} and photonic crystals \cite{lu15}. 

More generally, magnetic monopoles are ubiquitous in parameter spaces of adiabatic systems, 
as demonstrated by Berry \cite{berry84}. Perhaps most well-known is the canonical example of 
a single spin in a slowly changing external magnetic field, as described by the Zeeman 
interaction. Due to the spherical symmetry of this system, the monopole 
is forced to the origin of parameter space, where the external magnetic field vanishes. The 
corresponding magnetic charge is essentially the quantum number along the quantization axis 
of the instantaneous  spin eigenstate. Such synthetic magnetic monopoles in spin-like systems 
have been studied experimentally in a wide range of settings 
\cite{tomita86,bitter87,suter87,miniatura92,leek07,arai18}. 

Here, we examine magnetic monopoles in spin systems with broken spherical symmetry. 
Specifically, we provide a proof-of-concept demonstration of monopole field textures in 
the simplest nontrivial case consisting of a pair of interacting spins. Our purpose is to 
demonstrate a mechanism for how tunable magnetic monopole structures can be created 
in spin composites exposed to slowly varying external magnetic fields. 

Our system consists of two identical spin-$\frac{1}{2}$ particles in a slowly changing external 
magnetic field ${\bf b}$. The two spins ${\bf s}_1$ and ${\bf s}_2$ are coupled by a nonzero 
uniaxial exchange (Ising) interaction in the $z$ direction, combined with a Dzyaloshinskii-Moriya 
interaction (DMI) term. The Hamiltonian reads
\begin{equation}
H ({\bf b};J,{\bf D}) = {\bf b} \cdot {\bf S} + 4J s_1^z s_2^z + 
4{\bf D} \cdot \left( {\bf s}_1 \times {\bf s}_2 \right)  
\label{eq:ham}
\end{equation}
with ${\bf S} = {\bf s}_1 +  {\bf s}_1$ the total spin and $J$ the Ising coupling strength. 
The DMI vector ${\bf D}$ is confined to the $b_xb_z$ plane, i.e.,  ${\bf D} = D (\sin \vartheta,0,
\cos \vartheta)$ with $\vartheta$ the angle between the DMI and Ising axes \cite{remark1}. 
For notational convenience, we represent the spin-spin coupling by the vector $\vec{g} = 
(J,{\bf D})$. The system possesses cylindrical symmetry in the special case where 
$\vec{g} = (J,0,0,D)$; full rotation-symmetry is restored when $\vec{g} = \vec{0}$. In the 
general case with $J,D,\sin \vartheta \neq 0$, both these symmetries are broken.   

\begin{figure*}[htb]
\centering
\includegraphics[width=0.32\textwidth]{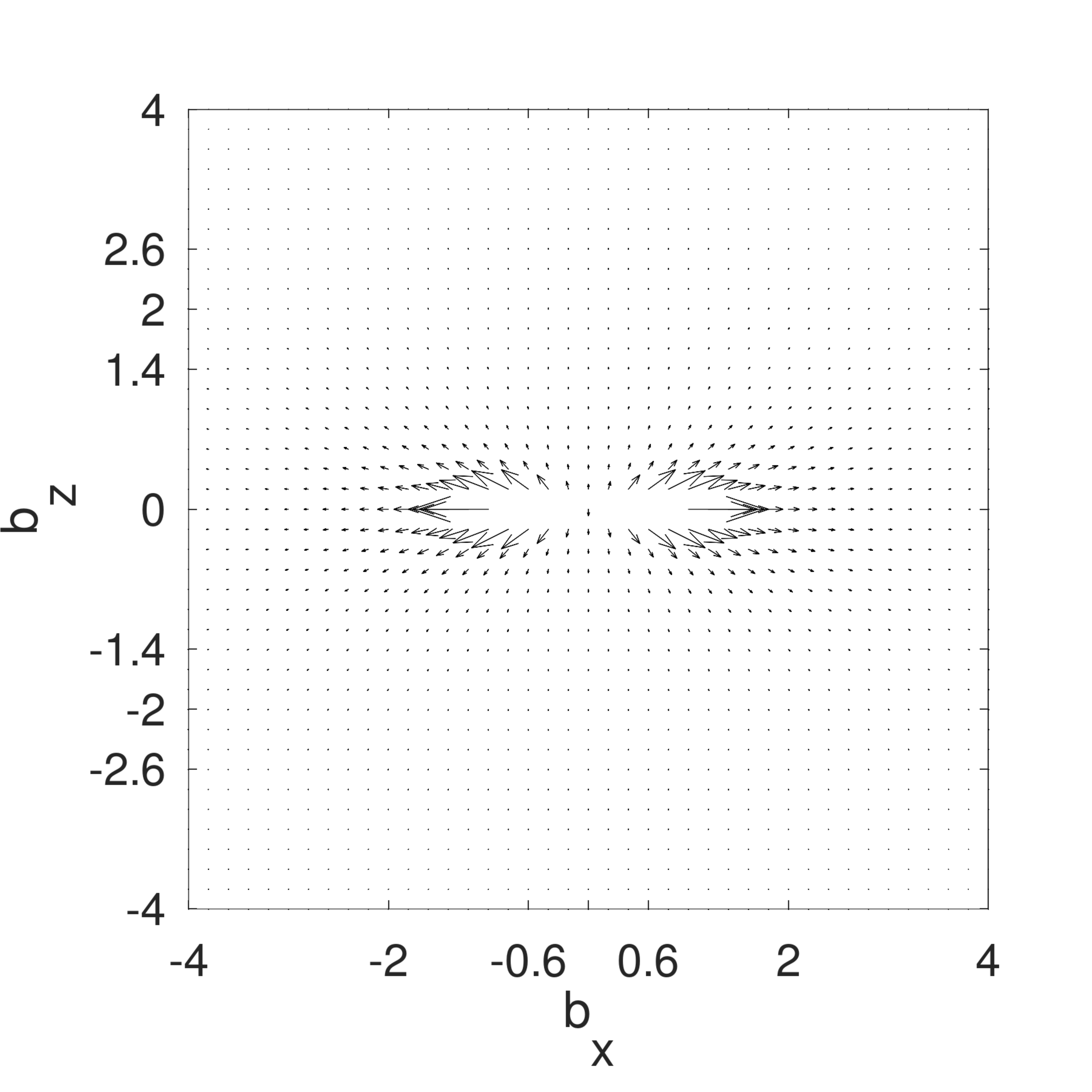}
\includegraphics[width=0.32\textwidth]{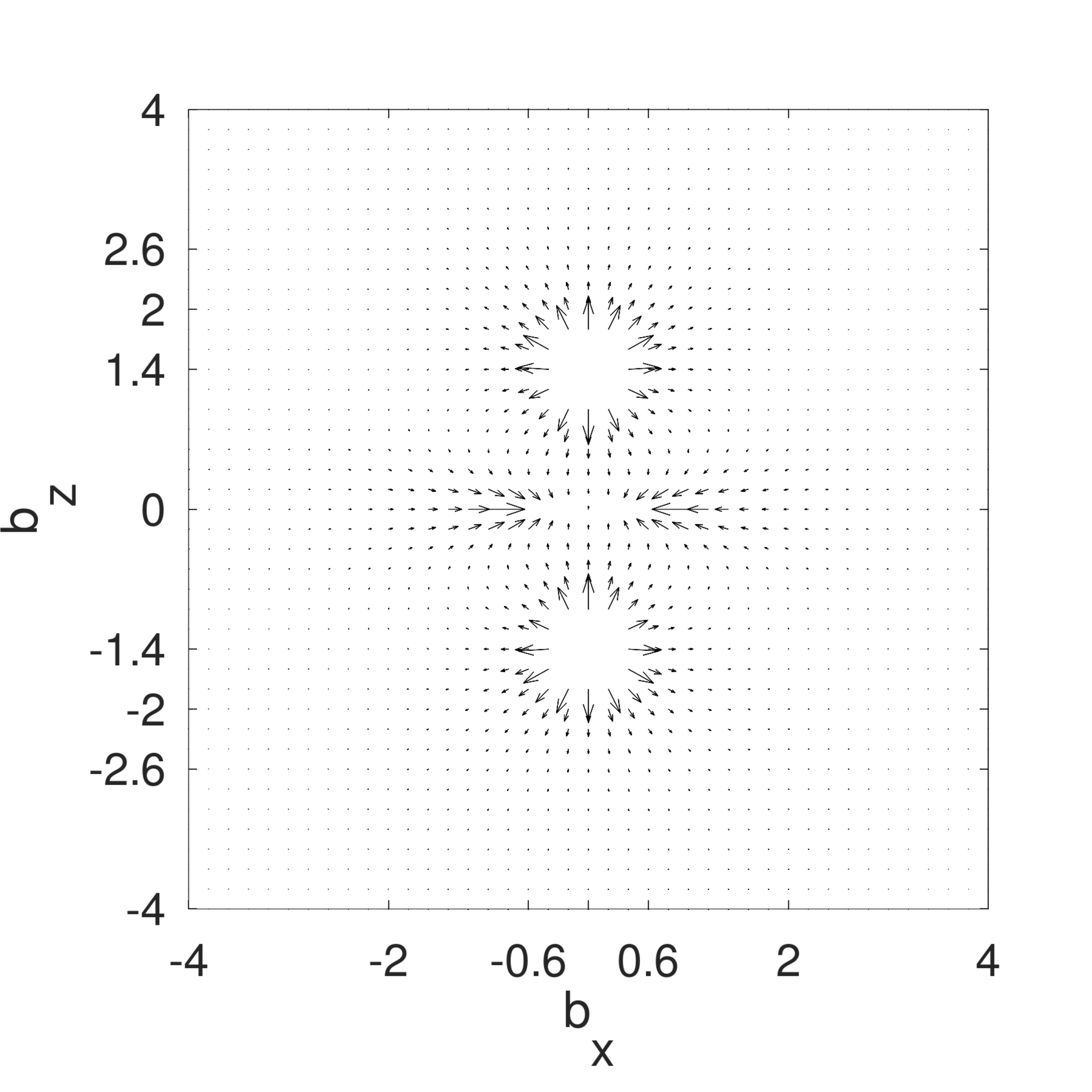}
\includegraphics[width=0.32\textwidth]{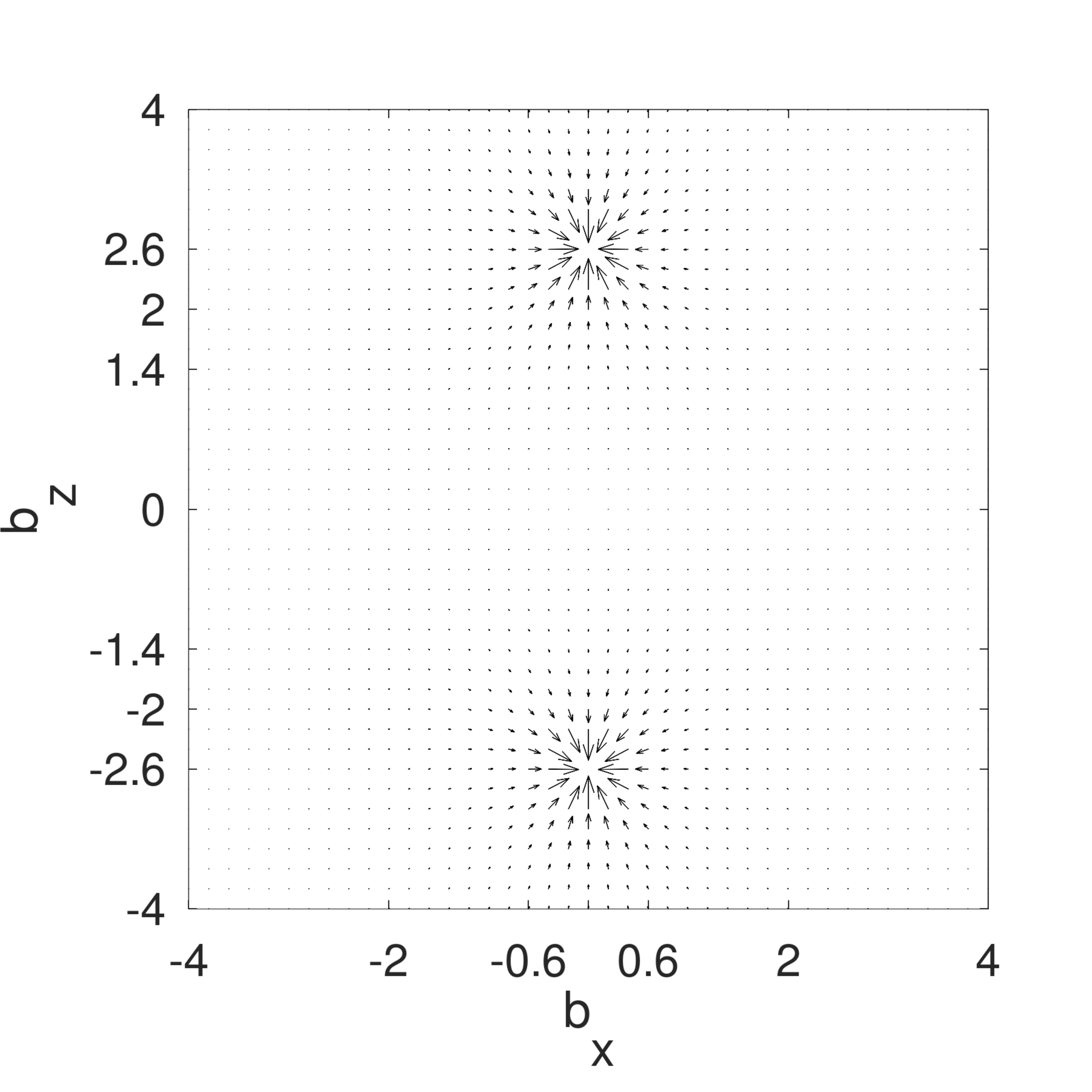}
\includegraphics[width=0.32\textwidth]{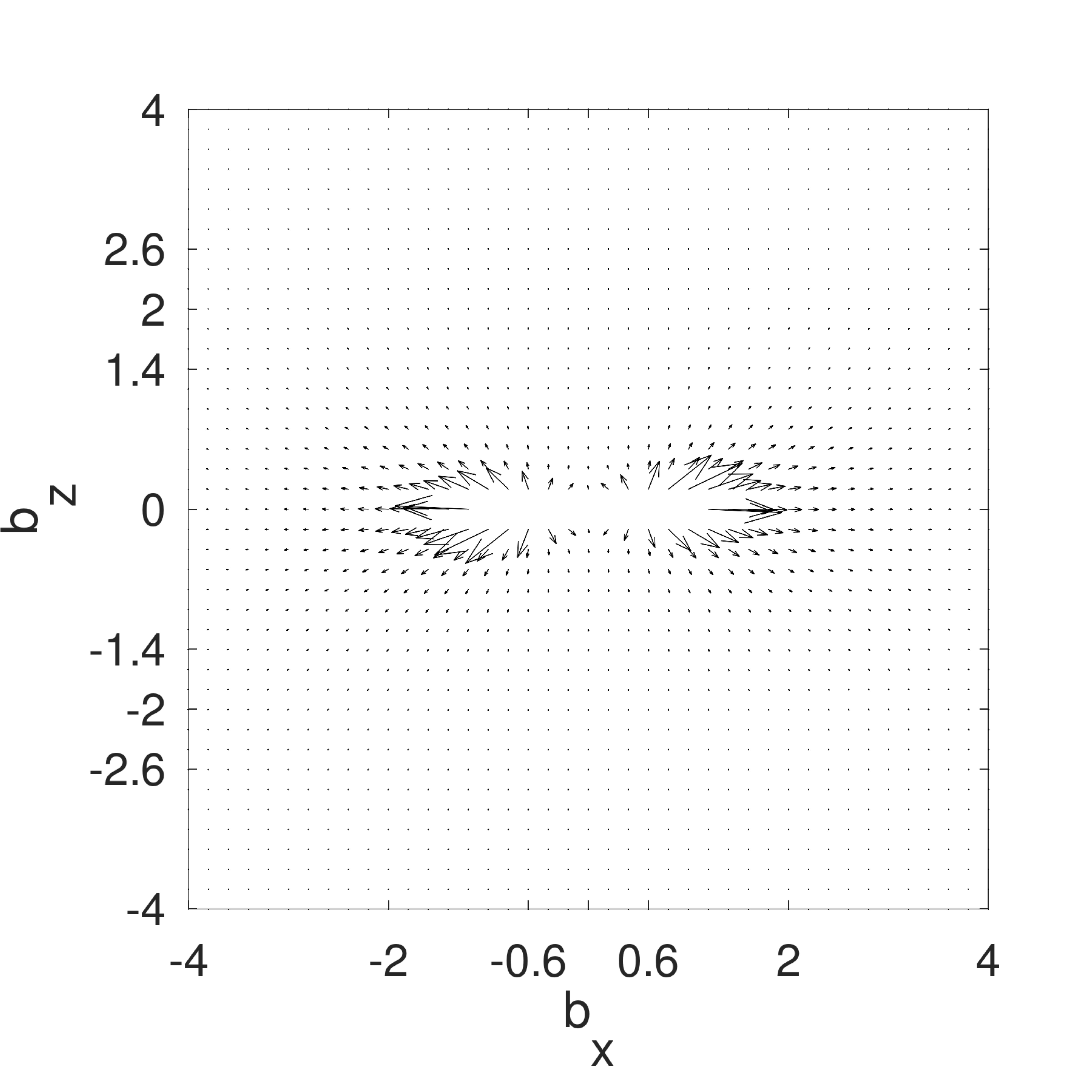}
\includegraphics[width=0.32\textwidth]{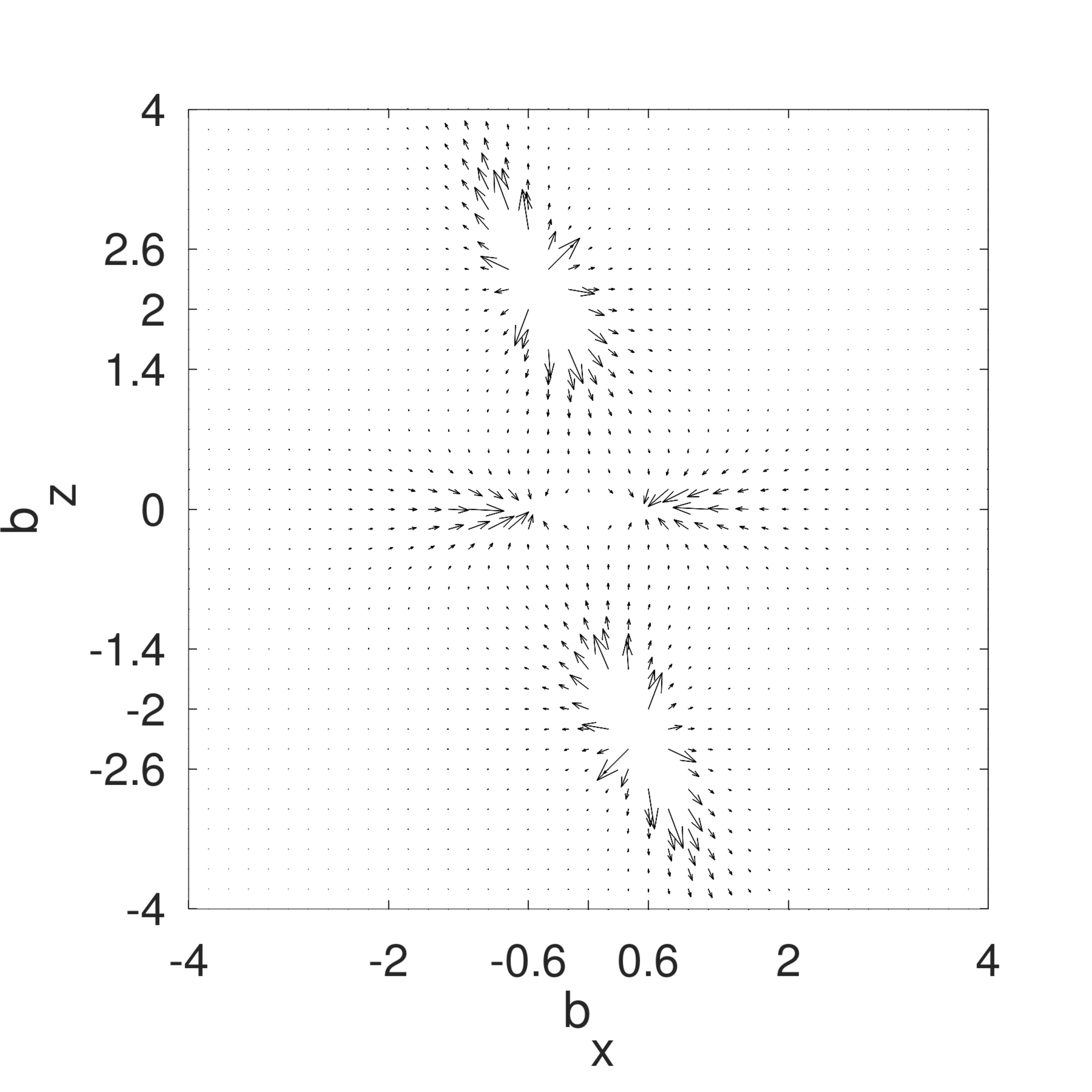}
\includegraphics[width=0.32\textwidth]{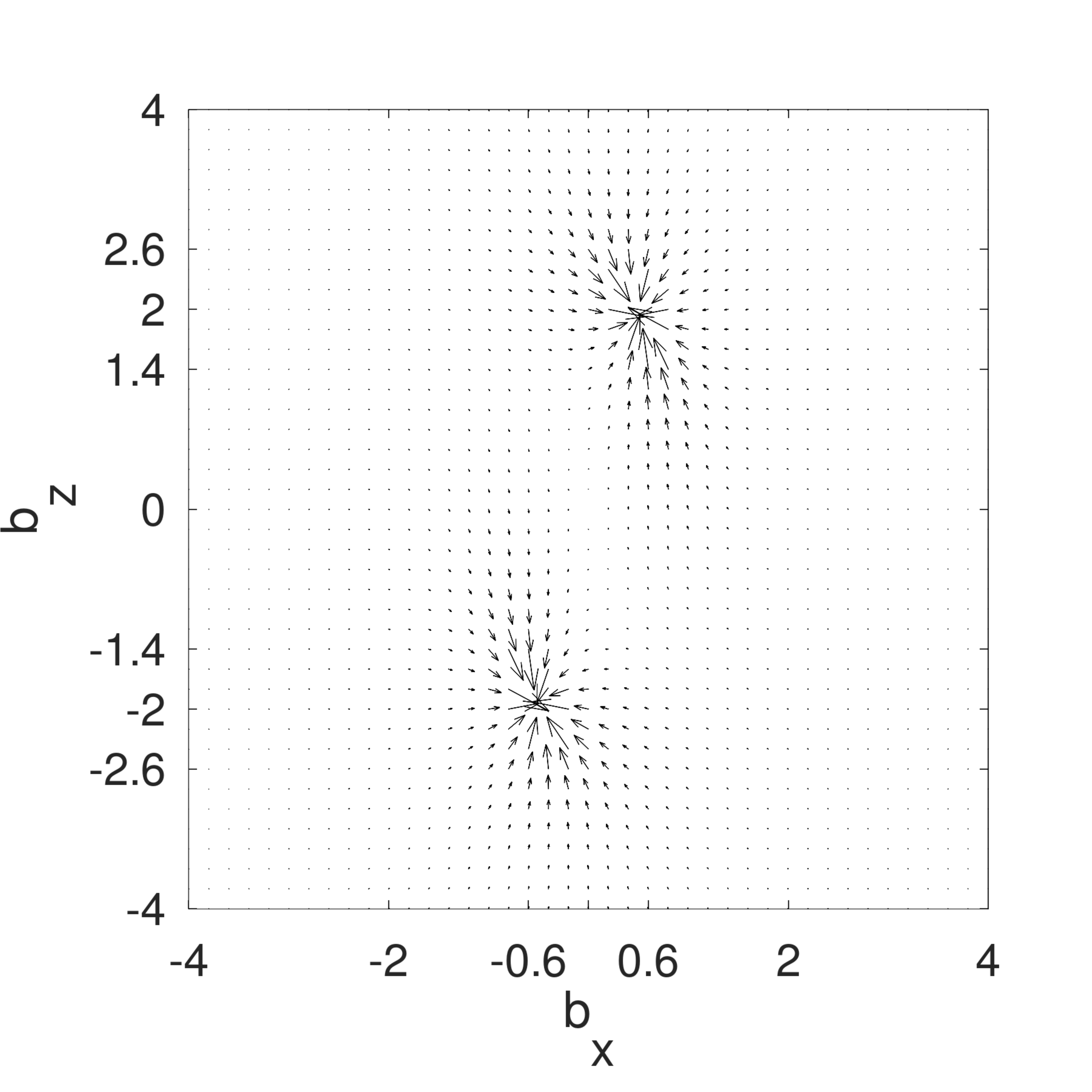}
\includegraphics[width=0.32\textwidth]{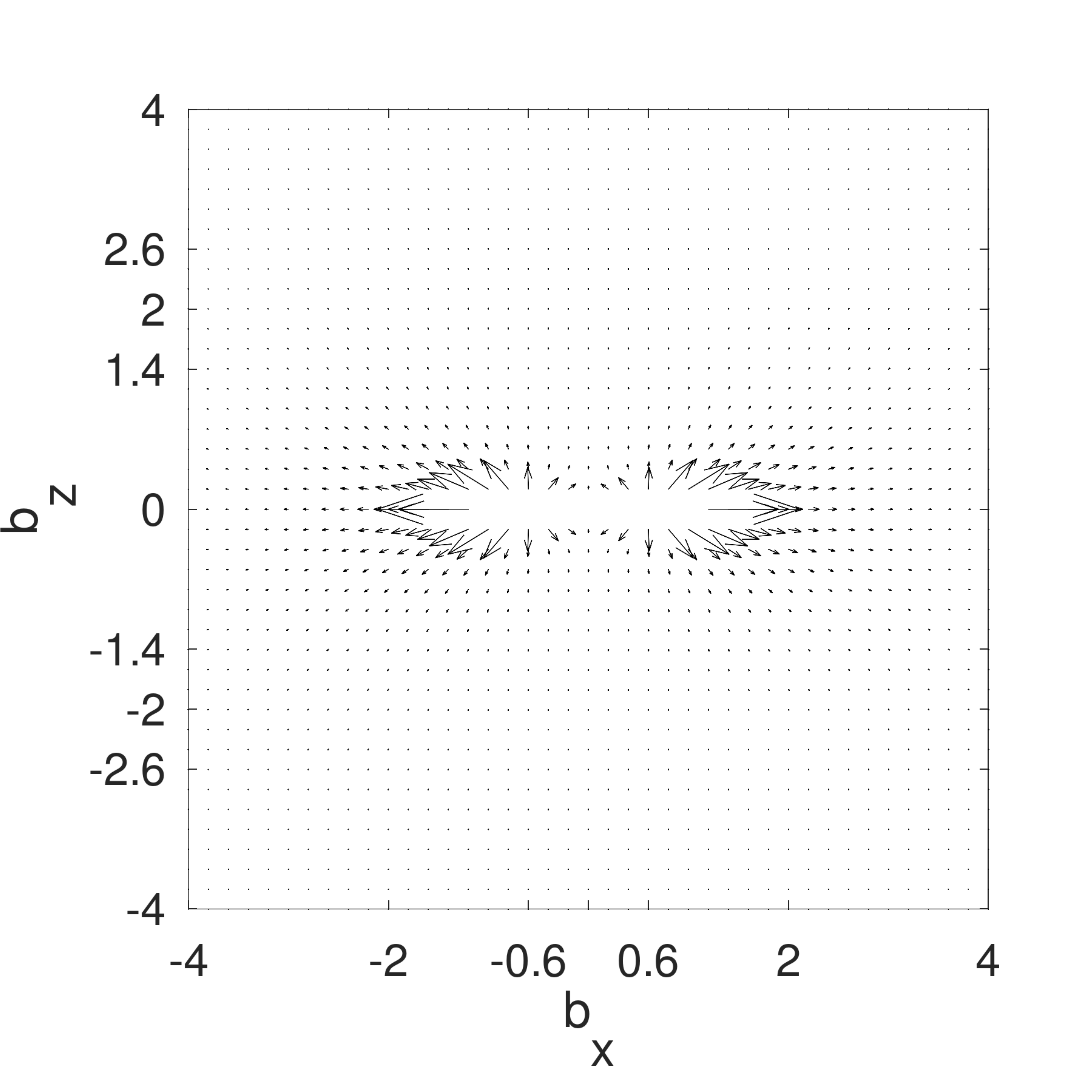}
\includegraphics[width=0.32\textwidth]{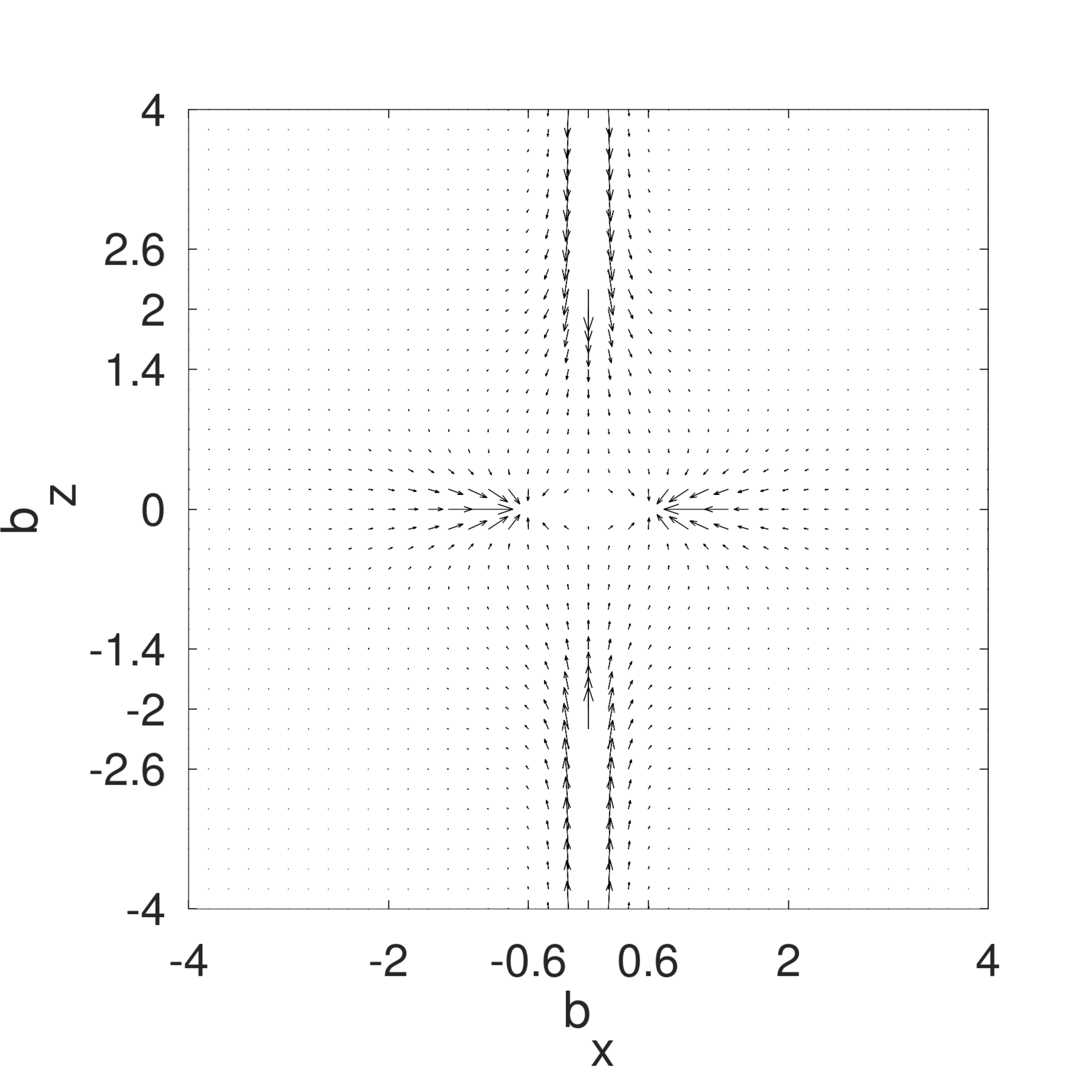}
\includegraphics[width=0.32\textwidth]{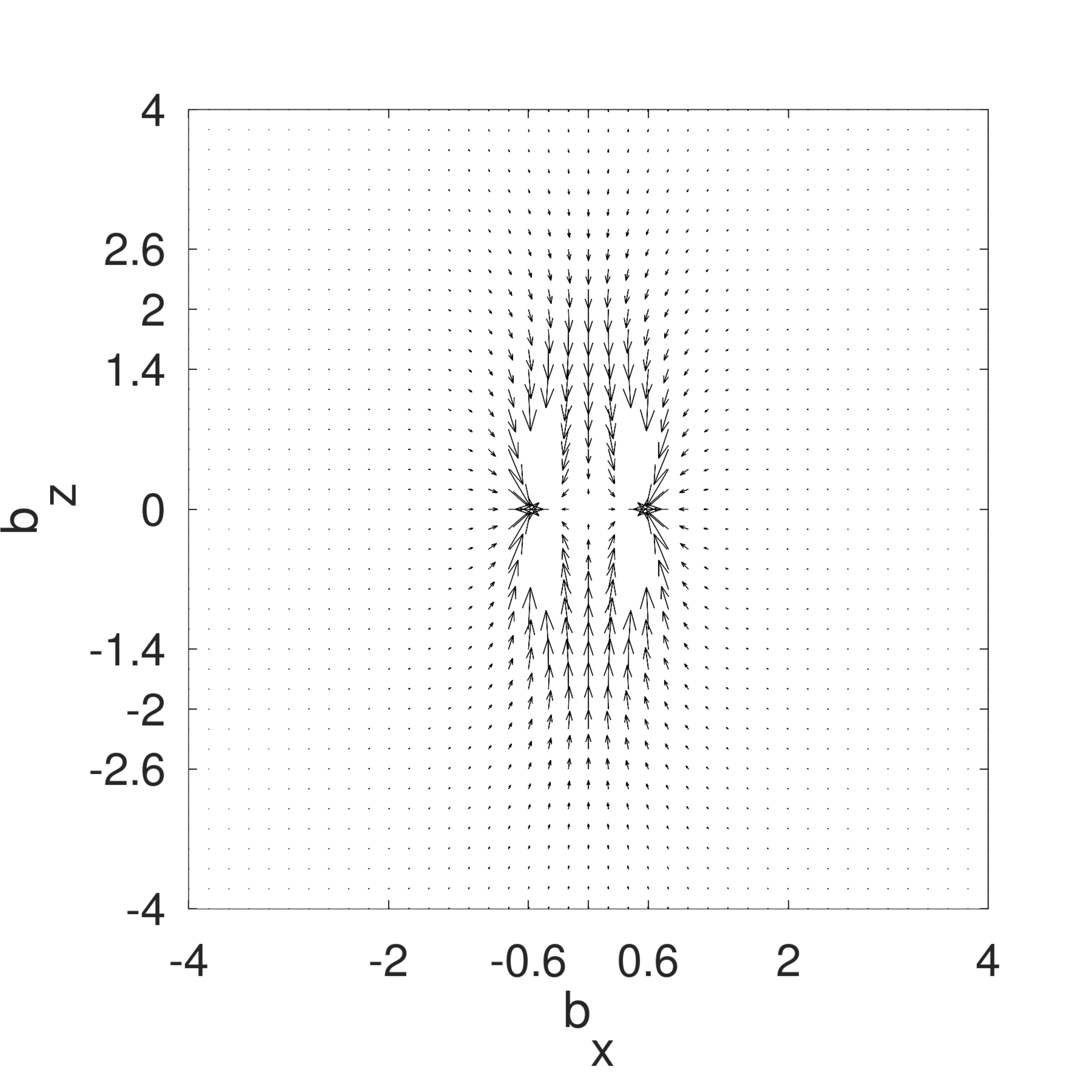}
\caption{Two-dimensional cuts of the synthetic magnetic fields for $J=1$ and $D=0.3$. 
From left to right in each row are shown ${\bf B}^{(1)}$, ${\bf B}^{(2)}$, and ${\bf B}^{(3)}$ 
with total magnetic charges $Q^{(1)} = +1$, $Q^{(2)} = 0$, and $Q^{(3)} = -1$, respectively. 
Upper row shows $\vartheta = 0^{\degree}$, where monopoles are expected at the energy 
crossing points that are found at $(0,0,0)$, $(0,0, \pm (2J +2D)) = (0,0,\pm 2.6)$, and 
$(0,0, \pm (2J -2D)) = (0,0,\pm 1.4)$. Middle row shows $\vartheta = 60^{\degree}$. Lower 
row shows $\vartheta = 90^{\degree}$ with all monopoles at $(\pm 2D,0,0) = (\pm 0.6, 0,0)$. 
Note that the there is a nonzero magnetic charge $q_{+}^{(2)} + q_{-}^{(2)} = -1$ associated 
with ${\bf B}^{(2)} (\vartheta = 90^{\degree})$ as two monopoles, each with charge 
$+\frac{1}{2}$,have been moved to infinity for this angle. Note that large arrows in the 
plot of ${\bf B}^{(2)} (\vartheta = 90^{\degree})$ have been omitted along the $b_z$ 
axis for magnetic field strengths $|b_z|>2$, in order to avoid obscuring the neighboring 
arrows.}
\label{fig:triplet}
\end{figure*}

\begin{figure*}[htb]
\centering
\includegraphics[width=0.32\textwidth]{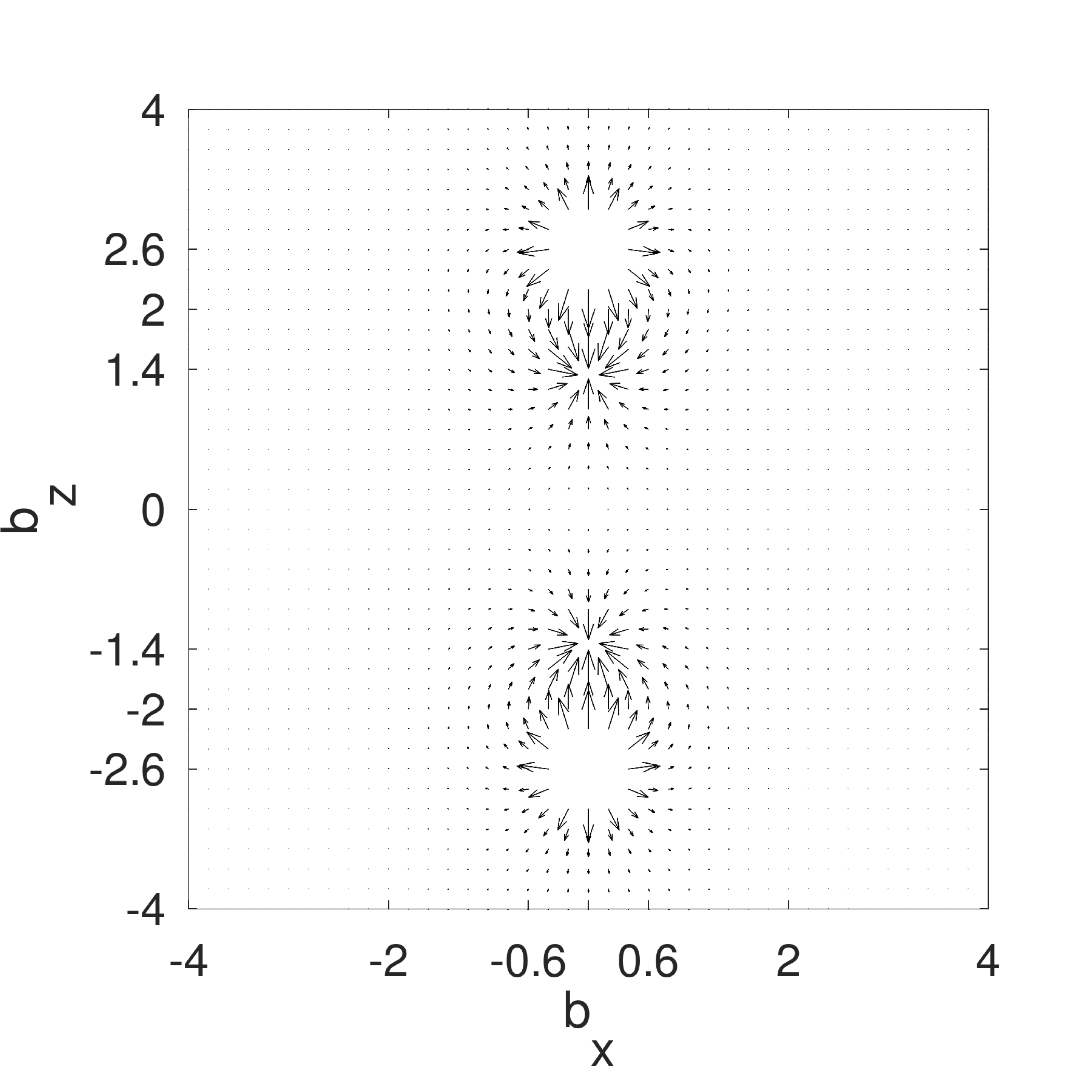}
\includegraphics[width=0.32\textwidth]{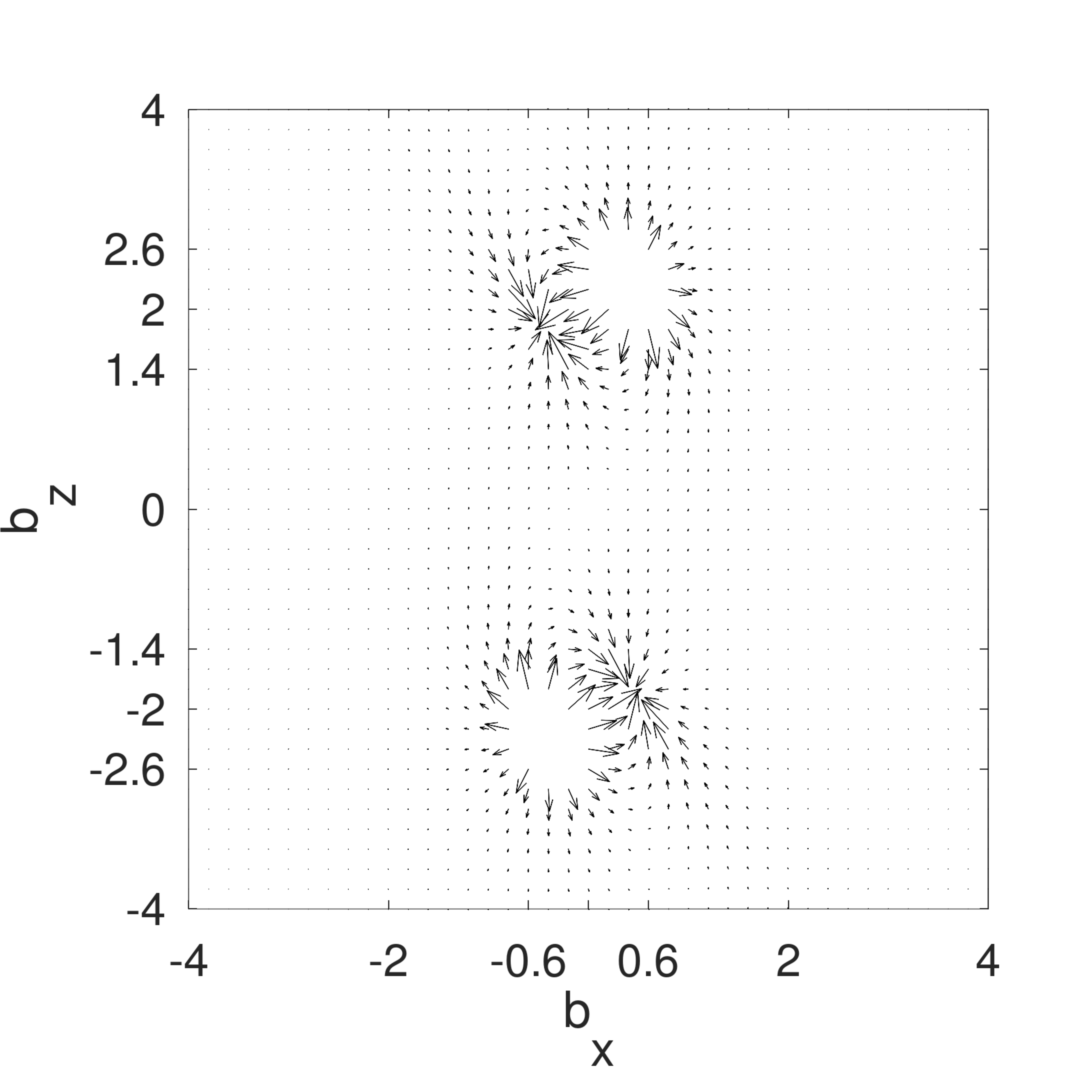}
\includegraphics[width=0.32\textwidth]{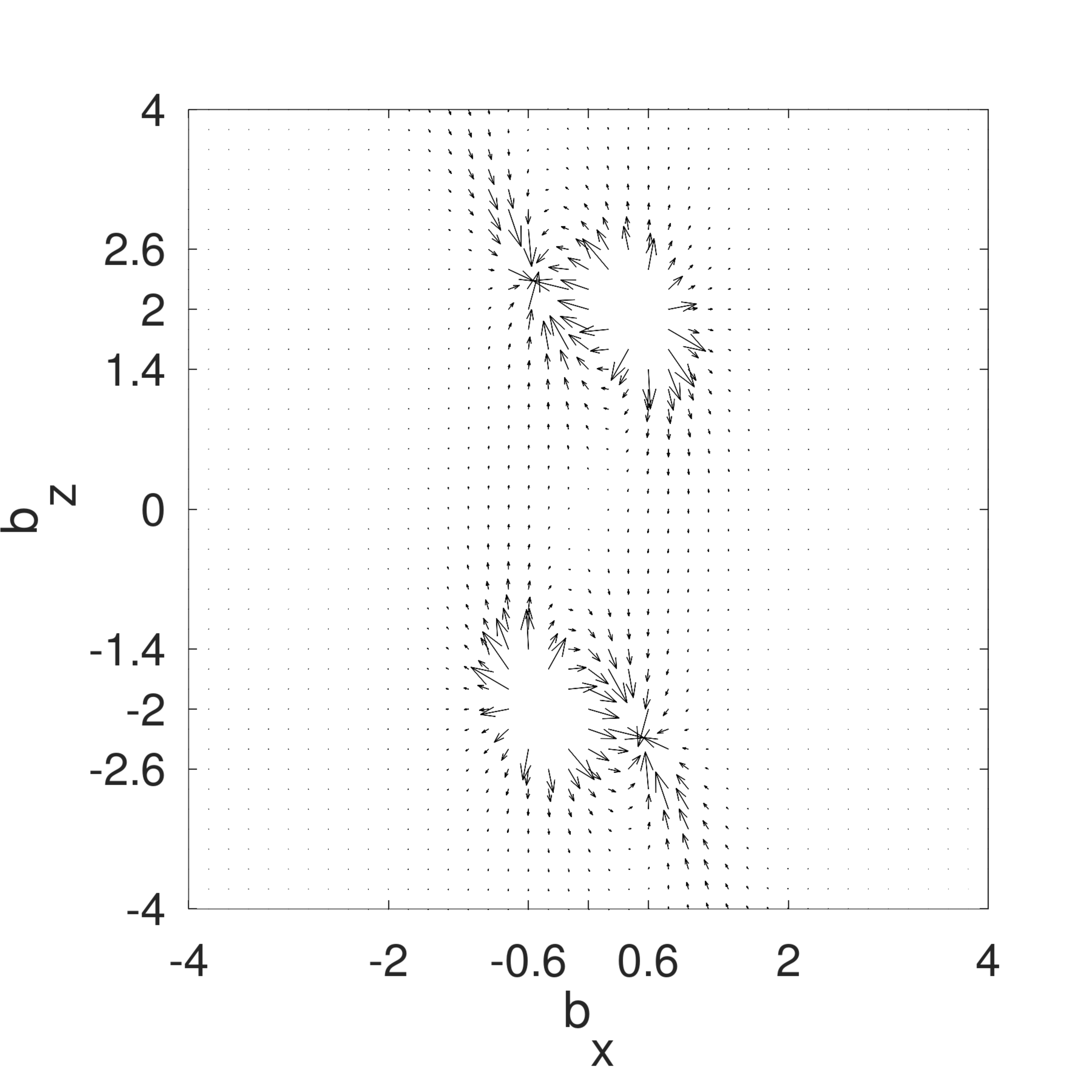}
\includegraphics[width=0.32\textwidth]{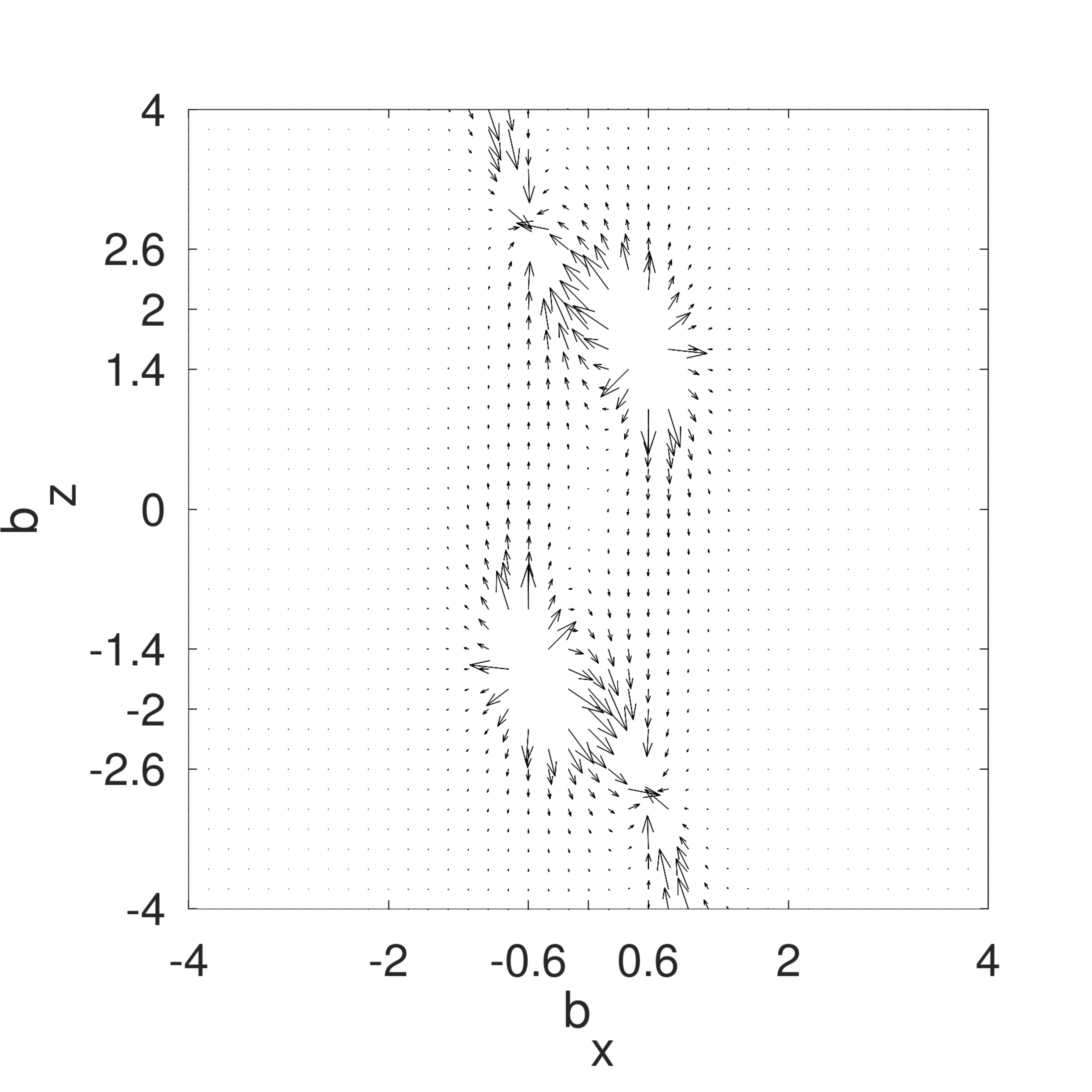}
\includegraphics[width=0.32\textwidth]{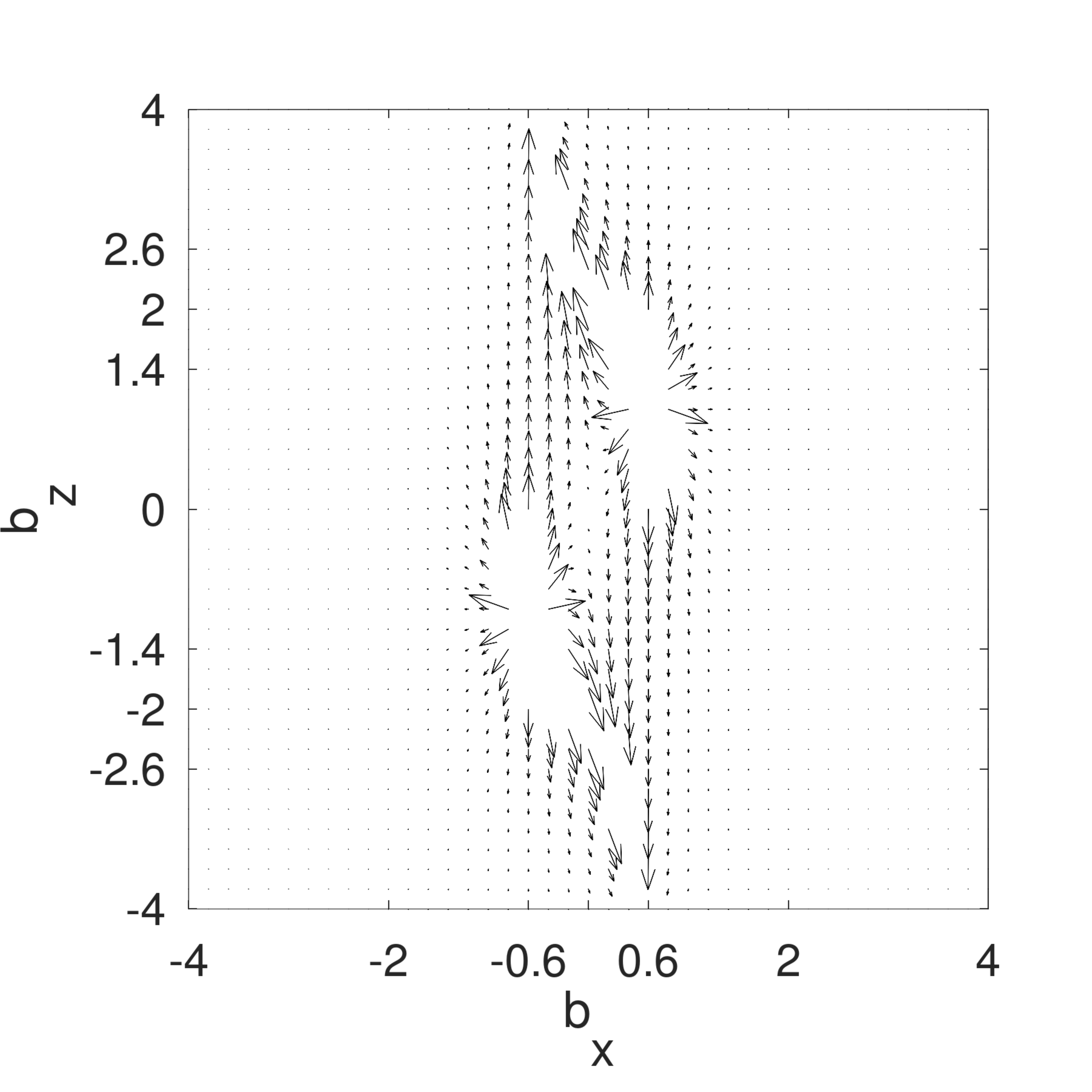}
\includegraphics[width=0.32\textwidth]{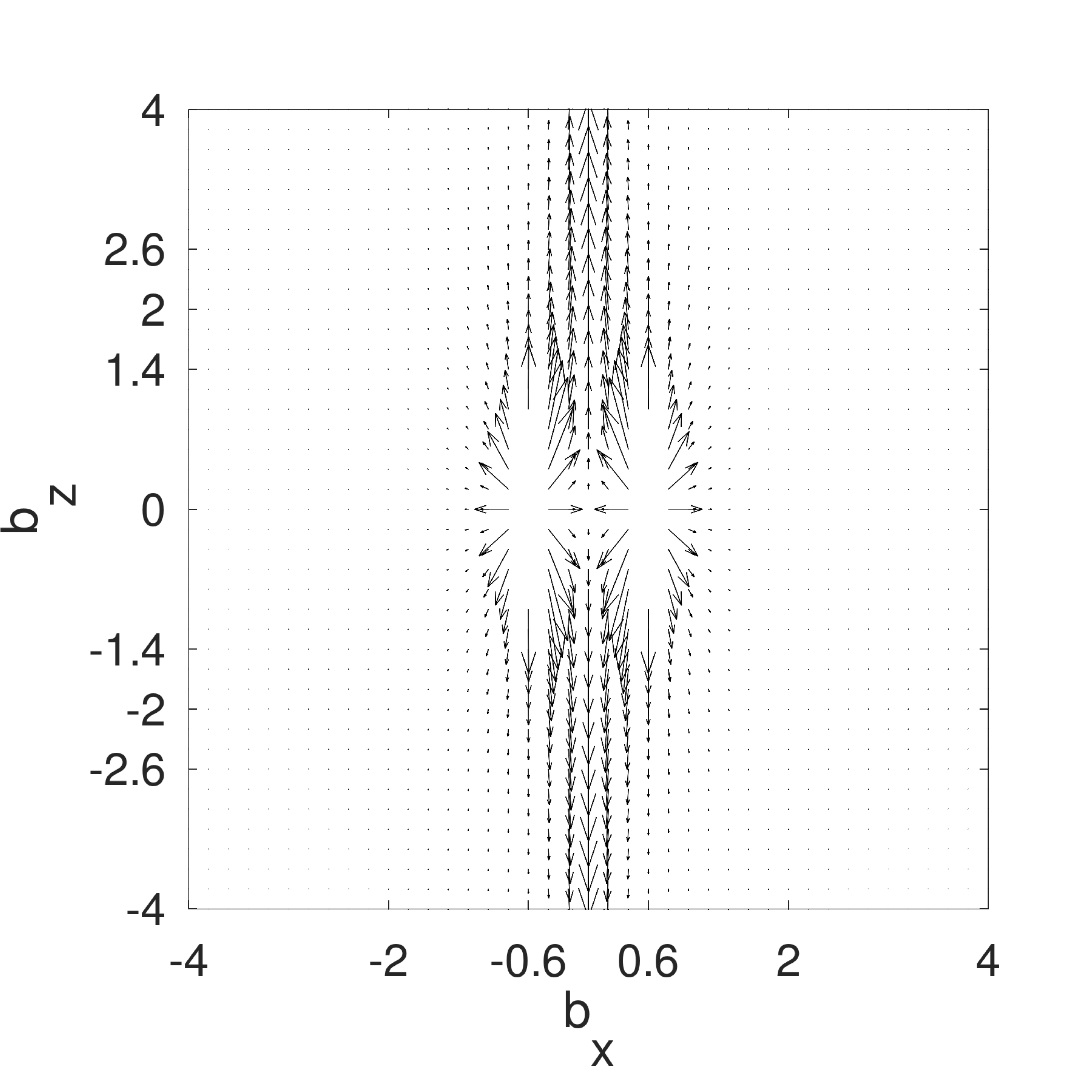}
\caption{Two-dimensional cuts of ${\bf B}^{(4)} ({\bf b}; J=1,D=0.3,\vartheta)$ for rotated DMI. 
From upper left to lower right: $\vartheta = 0^{\degree}, 45^{\degree}, 60^{\degree}, 
70^{\degree}, 80^{\degree}, 90^{\degree}$. The net magnetic charge vanishes for all 
$\vartheta \neq 90^{\degree}$. The progression to the net charge $q_{+}^{(4)} + q_{-}^{(4)} = +1$ 
by moving two monopoles, each with charge $-\frac{1}{2}$, to infinity, is visible.}
\label{fig:singlet}
\end{figure*}

In parameter space defined by the external field ${\bf b}$, there is a synthetic magnetic 
field \cite{berry84}  
\begin{eqnarray}
\boldsymbol{B}^{(k)} ({\bf b};\vec{g}) = i \sum_{l \neq k} 
\frac{{\bf S}_{kl} ({\bf b};\vec{g}) \times {\bf S}_{lk} ({\bf b};\vec{g})}{[E_l ({\bf b};\vec{g}) - 
E_k ({\bf b};\vec{g})]^2} 
\label{eq:fd}
\end{eqnarray}
associated with each energy eigenstate $\ket{\psi_k ({\bf b};\vec{g})}$, $k=1,\ldots ,4$, 
where ${\bf S}_{kl} ({\bf b};\vec{g}) = \bra{\psi_k ({\bf b};\vec{g})} {\bf S} 
\ket{\psi_l ({\bf b};\vec{g})}$. These fields define monopole charges $q_{\mu}^{(k)}$ 
that are located at energy crossing points ${\bf b}_{\mu}^{(k)}$. The charges are computed 
by means of Gauss' law 
\begin{eqnarray}
\frac{1}{4\pi} \iint_{\partial V} \boldsymbol{B}^{(k)} ({\bf b};\vec{g}) \cdot d{\bf S} = 
\sum_{\mu \in V} q_{\mu}^{(k)} , 
\label{eq:gauss}
\end{eqnarray}
where $V$ is a finite volume enclosed by a smooth orientable surface $\partial V$ in parameter 
space. 

The following sum rules are useful for analyzing the monopole distribution of the system. 
First, 
\begin{eqnarray}
\sum_k \sum_{\mu \in V} q_{\mu}^{(k)} = 0 
\label{eq:sum_rule_total_charge}
\end{eqnarray}
for any $V$, which can be seen by combining the identity \cite{{chruscinski04}} 
\begin{eqnarray}
\sum_k \boldsymbol{B}^{(k)} ({\bf b};\vec{g}) = 0 ,
\label{eq:sum_rule_B}
\end{eqnarray}
with Eq.~(\ref{eq:gauss}). Secondly, the total magnetic charge  
\begin{eqnarray}
Q^{(k)} = \sum_{\mu} q_{\mu}^{(k)} 
\label{eq:sum_rule_single_state}
\end{eqnarray}
for a given state $\psi_k$ is preserved under changes of $\vec{g}$. This follows from the 
observation that pairwise crossings of states occur at hypersurfaces of codimension $3$  
in the extended parameter space $({\bf b},J,{\bf D})$. According to the von Neumann-Wigner 
theorem \cite{vonneumann29}, this implies that independent variation of the three 
components of ${\bf b}$ is sufficient to induce point-like energy level crossings in the 
parameter space of the slowly changing magnetic field, no matter the form of spin-spin 
interaction. The total charge $Q^{(k)}$ is thus a topological invariant of the state $\psi_k$.  

We are now prepared to examine the synthetic magnetic fields in the spin-pair system. To 
determine the total magnetic charges, we first consider the pure Zeeman case ($J=D=0$). 
Here, the energy eigenstates coincide with the singlet-triplet states $\ket{S,M;{\bf n} \cdot 
{\bf S}}$, $S=0,1$, $M=-S,\ldots,S$, along the direction ${\bf n} = {\bf b} / |{\bf b}|$ of the 
instantaneous external magnetic field. As shown by Berry \cite{berry84}, one obtains 
${\bf B}^{(k)} ({\bf b};\vec{0}) \equiv {\bf B}^{(M)} ({\bf b};\vec{0}) = -M{\bf b}/|{\bf b}|^3$, 
which is a purely monopolar field. Thus, each synthetic magnetic field ${\bf B}^{(M)}$ 
correspond to a total magnetic monopole charge $Q^{(k)} \equiv Q^{(M)} =-M$ located at 
the origin of parameter space. 

Next, we add a nonzero Ising term, while keeping a vanishing DMI ($J \neq 0, D = 0$). 
The spin singlet remains decoupled from the triplet states, and its corresponding synthetic 
magnetic field therefore vanishes. Due to the cylindrical symmetry of the system, the energies 
are independent of the azimuthal spherical angle, which implies that the monopoles must 
be located on the $b_z$ axis \cite{remark2}. Indeed, by diagonalizing the Hamiltonian in the 
triplet subspace, one finds intersection points only at ${\bf b}_0^{(k)} = (0,0,0)$ and 
${\bf b}_{\pm}^{(k)} = (0,0,\pm 2J)$ in parameter space. The corresponding magnetic 
charges $q_0^{(k)}$ and $q_{\pm}^{(k)}$ can be found by numerically performing the 
integration in Eq.~(\ref{eq:gauss}) around each of  these intersection points. One finds 
\begin{eqnarray}
q_{0}^{(1)} & = & +1;
\nonumber \\ 
q_0^{(2)} & = & -1, \ \ q_{+}^{(2)} = q_{-}^{(2)} = +\frac{1}{2}; 
\nonumber \\ 
q_{+}^{(3)} & = & q_{-}^{(3)} =  -\frac{1}{2} , 
\end{eqnarray}
which confirm the sum rules in Eqs.~(\ref{eq:sum_rule_total_charge}) and 
(\ref{eq:sum_rule_single_state}). Note the appearance of half-integer magnetic charges for 
$\psi_2$ and $\psi_3$. 

We now turn to the general case where both $J$ and $D$ are nonvanishing. The broken 
symmetry caused by rotating ${\bf D}$ creates a nontrivial pattern of intersections of all 
four states, which in turn shows up as a nontrivial distribution of magnetic charges. Figure 
\ref{fig:triplet} shows the synthetic fields ${\bf B}^{(k)}$, $k=1,2,3$, each for $\vartheta = 
0^{\degree}, 60^{\degree},90^{\degree}$. For clarity, we show two-dimensional cuts of the 
field textures that contain the monopole charges. The remaining integer valued charges of 
$\psi_1$ and $\psi_2$ at the origin are now split into pairs of half-integer charges for 
$\vartheta \neq 0^{\degree}$. We note that no further splitting can take place by introducing 
other types of spin-spin interaction terms in the Hamiltonian, since the magnitude of each 
charge is $\frac{1}{2}$, which is the smallest allowed value  \cite{berry84}. The cylindrical 
symmetry still holds for $\vartheta = 0^{\degree}$, which forces the monopoles to remain 
on the $b_z$ axis in this case. For $\vartheta \neq 0^{\degree}$, the symmetry is lowered 
and the magnetic charges move into the $b_xb_z$ plane, but differently for $\psi_2$ and 
$\psi_3$, thereby creating a nonzero local total magnetic charge in the 
$\{ \psi_1, \psi_2,\psi_3 \}$ manifold. These nonzero local net charges are exactly 
cancelled by the local magnetic charges of $\psi_4$, as shown in Fig.~\ref{fig:singlet}, 
which confirms the sum rule in Eq.~(\ref{eq:sum_rule_total_charge}). 

By numerically computing the flux of ${\bf B}^{(4)}$ through a surface $\partial V$ that 
encloses the two magnetic charges at ${\bf b}_{\pm}^{(4)} = (\pm 2D,0,0)$ for 
$\vartheta = 90^{\degree}$, one finds the net magnetic charge $q_+^{(4)} + q_-^{(4)} = +1$, 
no matter the size of the region $V$ bounded by $\partial V$. Thus, a nonzero magnetic 
charge has been created. This is still consistent with the sum rule in 
Eq.~(\ref{eq:sum_rule_single_state}), as two monopoles, 
each with charge $-\frac{1}{2}$, have been moved to infinity when the DMI vector is 
rotated towards $\vartheta = 90^{\degree}$. To clearly show this progression, we have 
included the intermediate cases $\vartheta = 45^{\degree}, 70^{\degree}, 80^{\degree}$
in Fig.~\ref{fig:singlet}. As can be seen in Fig.~\ref{fig:triplet}, there is a nonzero net 
magnetic charge $q_+^{(2)} + q_-^{(2)} = -1$ for $\psi_2$ at $\vartheta = 90^{\degree}$. 
Just as for $\psi_4$, these charges are located at $(\pm 2D,0,0)$, which again confirms 
Eq.~(\ref{eq:sum_rule_total_charge}). 

Contrary to the Zeeman case, the synthetic magnetic field textures shown in Figs.~\ref{fig:triplet} 
and \ref{fig:singlet} are not purely monopolar in the presence of spin-spin interaction, i.e., 
it can be verified that 
\begin{eqnarray}
{\bf B}^{(k)} ({\bf b}; \vec{g} \neq \vec{0}) & \neq & 
\sum_{\mu} q_{\mu}^{(k)} \frac{{\bf b} - {\bf b}_{\mu}^{(k)}}{\left| {\bf b} - 
{\bf b}_{\mu}^{(k)} \right|^3} .  
\end{eqnarray}
Stated differently, the synthetic `electrical current density' defined via the 
Ampere-Maxwell-type equation ${\bf j}^{(k)} ({\bf b};\vec{g}) = \nabla \times {\bf B}^{(k)} 
({\bf b};\vec{g})$ is nonvanishing for $\vec{g} \neq \vec{0}$. 

By extending the system to more than two interacting spins, the total charges $Q^{(k)}$ 
become integer or half-odd integer valued depending on whether the system contains an 
even or odd number of spins, respectively. For such multi-spin systems, more complex 
point-like structures of an increasing number of monopoles are expected, as the total 
charges $Q^{(k)}$ can take increasingly larger values with the number of involved spins.  

In conclusion, we have provided a proof-of-concept demonstration of nontrivial magnetic 
monopole structures in a system of two interacting spin-$\frac{1}{2}$ particles. These monopoles 
can appear at points in parameter space where the external magnetic field is nonzero. 
This is in sharp contrast to the pure Zeeman case where monopoles only can appear 
at vanishing external magnetic field. We have shown that by increasing the complexity 
of the spin-spin interaction, the magnetic charges can be split into smaller entities. 
Furthermore, a nonzero magnetic charge can be created by tuning the spin-spin interaction 
for certain states. The synthetic magnetic field textures are nonmonopolar, which correspond  
to a nonvanishing synthetic electrical current density in parameter space. Our findings show 
that systems of interacting spins can give rise to highly nontrivial magnetic monopole 
structures. Trajectories  of particles composed of interacting spins and moving in 
spatially inhomogeneous external magnetic fields are sensitive to the Lorentz-type 
forces induced by the monopoles and synthetic electric currents. This feature provides 
a tool for studying the synthetic magnetic field structure experimentally.

\section*{Acknowledgments}
E.S. acknowledges support from the Swedish Research Council (VR) under Grant No. 
2017-03832.


\begin{thebibliography}{99}
\bibitem{castelnovo08} C. Castelnovo, R. Moessner, and S. L.~Sondhi,  
Magnetic monopoles in spin ice, 
Nature {\bf 451}, 42 (2008). 
\bibitem{ray15} M. W. Ray, E. Ruokokoski, K. Tiurev, M. M\"ott\"onen, and D. S. Hall,
Observation of isolated monopoles in a quantum field, 
Science {\bf 348}, 544 (2015). 
\bibitem{fang03} Z. Fang, N. Nagaosa, K. S. Takahashi, A. Asamitsu, R. Mathieu, 
T. Ogasawara, H. Yamada, M. Kawasaki, Y. Tokura, and K. Terakura, 
The Anomalous Hall Effect and Magnetic Monopoles in Momentum Space, 
Science {\bf 302}, 92 (2003). 
\bibitem{lv15} B. Q. Lv, H. M. Weng, B. B. Fu, X. P. Wang, H. Miao, J. Ma, P. Richard, 
X. C. Huang, L. X. Zhao, G. F. Chen, Z. Fang, X. Dai, T. Qian, and H. Ding, 
Experimental Discovery of Weyl Semimetal TaAs, 
Phys. Rev. X {\bf 5}, 031013 (2015). 
\bibitem{lu15} L. Lu,, Z. Wang, D. Ye, L. Ran, L. Fu, J. D. Joannopoulos, and M. Soljaci\'{c}, 
Experimental observation of Weyl points, 
Science {\bf 349}, 622 (2015). 
\bibitem{berry84} M. V. Berry,
Quantal Phase Factors Accompanying Adiabatic Changes, 
Proc. R. Soc. London Ser. A {\bf 392}, 45 (1984).
\bibitem{tomita86} A. Tomita and R. Y. Chiao, 
Observation of Berry's Topological Phase by Use of an Optical Fiber, 
Phys. Rev. Lett. {\bf 57}, 937 (1986).
\bibitem{bitter87} T. Bitter and D. Dubbers, 
Manifestation of Berry's topological phase in neutron spin rotation, 
Phys. Rev. Lett. {\bf 59}, 251 (1987).  
\bibitem{suter87} D. Suter, G. C. Chingas, R. A. Harris, and A. Pines, 
Berry's phase in magnetic resonance, 
Mol. Phys. {\bf 61}, 1327 (1987). 
\bibitem{miniatura92} Ch. Miniatura, J. Robert, O. Gorceix, V. Lorent, S. Le Boiteux, 
J. Reinhardt, and J. Baudon, 
Atomic interferences and the topological phase, 
Phys.~Rev.~Lett.~{\bf 69}, 261 (1992). 
\bibitem{leek07} P. J. Leek, J. M. Fink, A. Blais, R. Bianchetti, M. G\"oppl, J. M. Gambetta, 
D. I. Schuster, L. Frunzio, R. J. Schoelkopf, and A. Wallraff, 
Observation of Berry's Phase in a Solid-State Qubit, 
Science {\bf 318}, 1889 (2007). 
\bibitem{arai18} K. Arai, J. Lee, C. Belthangady, D. R. Glenn, H. Zhang, and R. L. Walsworth, 
Geometric phase magnetometry using a solid-state spin,  
Nature Comm. {\bf 9}, 4996 (2018). 
\bibitem{remark1} The combination of Ising and DMI terms is chosen so as to give rise to 
a nontrivial but simply analyzable pattern of magnetic monopoles. One may envisage other 
combinations of coupling terms that can achieve different forms of monopole structures.  
\bibitem{chruscinski04} D. Chruscinski and A. Jamiolkowski, 
{\it Geometric Phases in Classical and Quantum Mechanics}, 
(Birkh\"auser, Basel, 2004). 
\bibitem{vonneumann29} J. von Neumann and E. P. Wigner, 
\"{U}ber das Verhalten von Eigenwerten bei adiabatischen Prozessen, 
Phys. Z. {\bf 30}, 467 (1929). 
\bibitem{remark2} Otherwise they would, as a consequence of the cylindrical symmetry, 
form ring-like structures around the $b_z$ axis, which is not allowed by the von Neumann-Wigner 
theorem. 
\end{thebibliography}
\end{document}